\documentclass[prl,twocolumn,showpacs]{revtex4-1}
\usepackage{amssymb,amsmath,bm,epsfig,graphicx,subfigure}
\usepackage{amssymb,amsmath,bm,epsfig,graphicx,subfigure}

\begin{document}

\title{Bending Rules for Nano-Kirigami}
\author{Bastien F. Grosso$^{1,3}$ and E. J. Mele$^{2,3}$}
    \email{mele@physics.upenn.edu}
    \affiliation{$^1$Institute of Theoretical Physics Ecole Polytechnique Federale de Lausanne (EPFL)  Lausanne CH-1015  \\
    $^2$Department of Physics and Astronomy  University of Pennsylvania Philadelphia PA 19104  USA\\
    $^3$Department of Physics Loughborough University LE11 3TU UK}
\date{\today}

\begin{abstract}
We combine large-scale atomistic modelling with continuum elastic theory to study the shapes of graphene
sheets embedding nanoscale kirigami. Lattice segments are selectively removed from a flat graphene sheet and the structure is allowed to close and reconstruct by relaxing in the third dimension. The surface relaxation is limited by a nonzero bending modulus which produces a smoothly  modulated landscape instead of the ridge-and-plateau motif found in macroscopic lattice kirigami. The resulting surface shapes and their interactions are well described by a new set of microscopic kirigami rules that resolve the competition between the bending and stretching energies.
\end{abstract}

\pacs{61.48.Gh,61.72.-y,46.70.Hg}
\maketitle

\medskip
Folding a two dimensional material lifts it into the third dimension enabling different physical
 functionalities. In a familiar example, folds can be introduced into a piece of paper to change its three dimensional shape with or without  allowing for tears. A lattice  model for the former case (kirigami) has been studied recently \cite{kirigami} demonstrating rules for generating three dimensional shapes by the selective removal of segments from a parent honeycomb lattice and closing the tears by folding. Because the folding rules so defined are essentially geometrical it is possible that they could find applications in two dimensional nanoscale materials and possibly even affect their electronic behavior \cite{YazyevReview}.

In this Letter we examine this possibility by combining large scale atomistic modelling \cite{lammps,fasolino} with analysis developed from long wavelength elastic theory \cite{LandauLifshitz,NelsonPeliti,SeungNelson} using a graphene sheet as a prototype. The models we adopt violate two central tenets of macroscopic lattice kirigami: (a) the bending modulus is nonzero prohibiting the formation of sharply folded edges and (b) the medium is compressible allowing the system to store energy in  shear and compressive strains. Thus, and perhaps not surprisingly, one finds that in a nanomaterial the regular faceted structures of macroscopic kirigami inevitably relax to softly rolling landscapes evocative of the English countryside. We find that these latter shapes, identified here in numerical calculations, can be accurately modelled and predicted using a new set of {\it microscopic}-kirigami rules appropriate in the weakly compressible limit.

The shapes shown in Figure 1 compare two representative kirigami-folded structures with their atomically relaxed counterparts. The top panels contain defects in which atoms are removed from a strip and the gap is eliminated by rejoining lattice sites along a line that terminates on compensating edge dislocations containing nearest neighbor 5- and 7- membered rings.  Fig. 1(c-d) illustrate a deflection of this structure into the third dimensions via sharp folds that vertically displace the left and right hand regions in the same (panel (c)) or opposite (d) directions. The defect energy density in this structure is confined to the edges of the folds so that the ``up-up" ({\it uu}) and ``up-down" ({\it ud}) patterns are degenerate \cite{kirigami}. Starting from these structures we minimized the structural energy of a variety of atomistic models using interaction potentials for carbon developed by Los and Fasolino (LF) \cite{lammps,fasolino} which allow bonds to rupture and reform and provide a useful description of the elastic properties for carbon derived materials in diverse bonding environments. The structures we develop should be contrasted with patterned graphenes containing large open perforations designed to allow reversible large amplitude deformations under mechanical loading \cite{graphenekirigamiBU,graphenekirigamiCornell}. They are more akin to the fully bonded defect structures containing height modulations found on scars that terminate on dislocation cores in some single layer graphenes produced by chemical vapor deposition \cite{lehtinen}. Two generic features of the fully relaxed structures are apparent in the lower panels of Figure 1(e-f). For both defects we find a smooth variation in elevation that persists into the far field with soft pleats sourced by their near field defect structures.

\begin{figure}
\begin{center}
\includegraphics[angle=0,bb= 0 0 310 257,width=\columnwidth]{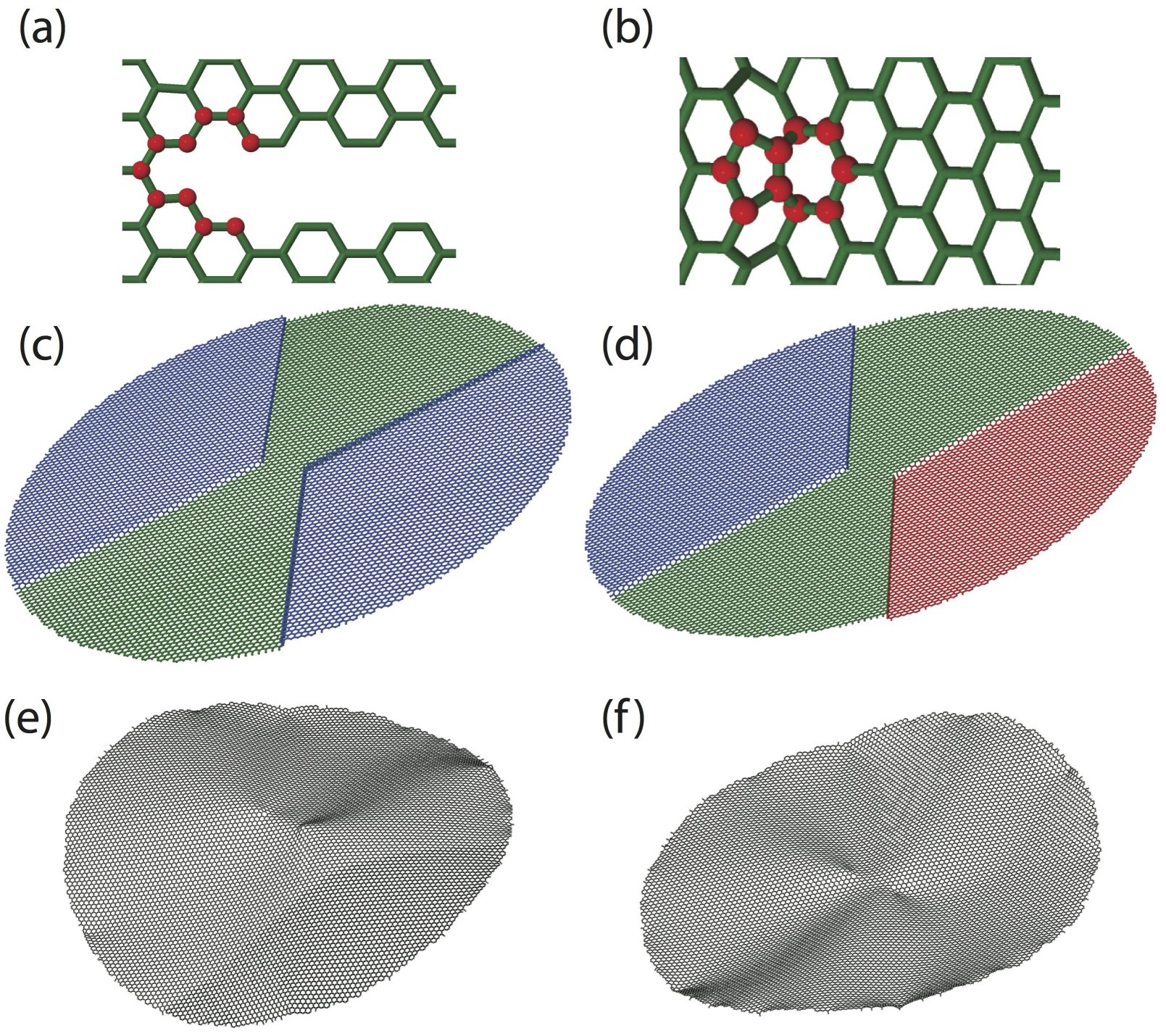}
\caption{\label{structures} Relaxation of a graphene kirigami in which atoms are removed from finite strip of flat graphene sheet (a) and the gap is rejoined, terminating on nearest neighbor pairs of 5-7 disclinations (b). In (c) and (d) this structure is folded into three dimensions following the rules for macroscopic lattice kirigami where sharp folds separate plateaus that are displaced out of the plane in the same direction {\it uu} (c) or in opposite directions {\it ud} (d). This microscopic structure relaxes to the shapes (e) and (f) generating a softly pleated landscape.}
\end{center}
\end{figure}

We quantify these observations by decomposing the height field $h(\textbf{r})$ on a disk of radius $R$ into angular harmonics
\begin{eqnarray}
h(\textbf{r}) = \sum_m \, h_m(r)e^{i m \phi}
\end{eqnarray}
Figure 2(a) shows the radial dependence $h_m(r)$ for the allowed even $m$ amplitudes in the shape in Figure 1(e). The relaxed structure is smooth, suppressing weight in its large $m$ modes and confining its amplitude to the $m=0, \pm 2$ deformations of the disk where $h_2(r)$ (Fig. 2(b)) is an increasing function of $r$ out to the boundary. The bending energy has an areal energy density $u_b= \kappa_b (\nabla^2 h)^2/2$  and it is extremized by solutions of the biharmonic equation $\nabla^4 h=0$. We find that the radial dependences of our relaxed structures $h_m(r)$ are quite well described by linear combinations of these solutions projected into each angular harmonic subspace. For example in the $m=2$ subspace the representation
\begin{eqnarray}
h_2(r)= h_{-2}(r) = a_2 + \frac{b_2}{r^2} + c_2 r^2 + d_2 r^4
\end{eqnarray}
describes the shape as shown in Figure 2(c). Truncating the expansion (1) to include only the $m=0$ and $m=\pm 2$ solutions provides an excellent reconstruction of the exact shape as demonstrated in Figure 3(a). The {\it ud} structure (Fig. 1(f)) similarly relaxes to a smooth landscape well described by a superposition $m=\pm 1, \pm 3$ angular harmonics.

\begin{figure}
\begin{center}
\includegraphics[angle=0,bb= 0 0 800 650,width=\columnwidth]{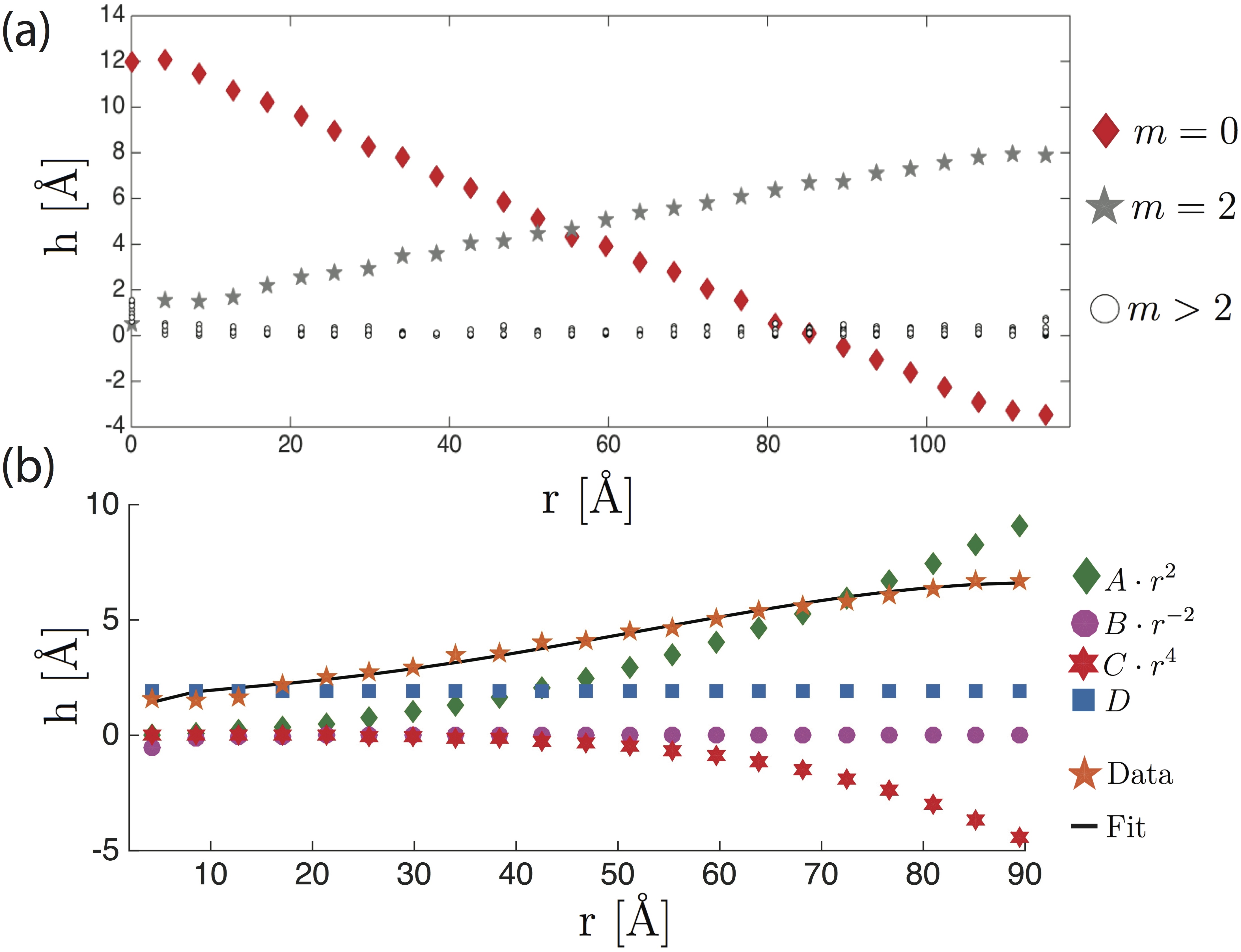}
\caption{\label{fits} (a) The height field for the relaxed {\it uu} is decomposed into its angular harmonics showing the radial dependence $h_m(r)$ of its dominant contributions from $m=0, \pm 2$. (b) The $m=2$ radial dependence is well described by four solutions of a biharmonic equation projected into the $m=2$ subspace.  The fit requires growing solutions with opposite signs which dominate the deflection in the far field.}
\end{center}
\end{figure}

 Note that the biharmonic equation admits {\it two} solutions that grow in the far field and generically these are both present in the relaxed structures but they always appear with opposite signs. Although it is tempting to attribute this to a boundary condition enforced at the edge of the disk, we find instead that this can be more easily understood as a global constraint on the shape. The growing solutions must compete in order to avoid a large strain energy penalty induced by their (locally) nonzero Gaussian curvatures. Note that a linear combination of the growing solutions in Eqn. 2 make a contribution to the Gaussian curvature that is {\it bilinear} in the expansion coefficients for $h_2$; explicitly we have for the determinant of the curvature tensor in the far field
\begin{eqnarray}
{\cal C}_2^> = -4 \left( c_2^2 + 6c_2 d_2 r^2 + 6 d_2^2 r^4 \sin^2(2 \phi) \right)
\end{eqnarray}
Following Nelson and Peliti \cite{NelsonPeliti} we recall that a coupling of the local Gaussian curvature to in-plane strain mediates nonlocal ultra-long range interactions between remote Gaussian curvatures, diverging in Fourier space $\propto q^{-4}$.  Consequently, for a large system under open boundary conditions we can avoid a macroscopic energy that grows faster than the system size if its integrated Gaussian curvature vanishes.  In the space of $m$-projected biharmonic solutions the residual Gaussian curvature cannot be made to vanish everywhere and with zero mean the residual curvature can be usefully described by its nonvanishing moments. For $m=2$ and using Eqn. 3 we find that the disk-integrated curvature vanishes if the boundary ratio $\nu = d_2 R^2/c_2 = - 0.423$, in good agreement with the ratio ($\sim -0.47$) obtained from our numerical calculations. We carried out similar analysis for different structures and in various angular momentum channels $m$ in the expansion (1) and find that the boundary ratio is $m$-dependent and consistent with our simulation data.

 The surfaces shown in Figure 1(e,f) are therefore determined by three rules that resolve the competition between its bending and stretching energy in the elastically stiff (weakly compressible) limit: (1) the height field smooths by relaxing its amplitude to its low order symmetry-allowed angular harmonics, (2) the radial dependence in each $m$-channel superposes biharmonic solutions thereby producing a low bending energy, (3) these appear in ``well-tempered" combinations that also avoid a large strain energy penalty by quenching the {\it integrated} Gaussian curvature.  The defect energy is then determined by the core energy of the terminal dislocations,  the bending energy in the extremal solution and the strain energy imposed by its residual Gaussian curvature.

 The argument given above fixes the amplitude ratio of the far field growing solutions but not their overall magnitudes which determines the degree of ``warping" of the kirigami-ed disk. A scaling argument reveals that the latter is determined by a boundary energy on the perimeter of the disc, presumably arising from the inequivalence of bulk (area) and surface (perimeter) interactions. For example, a structure with $c_2 \neq 0$ that results from a boundary interaction proportional to $R$ and is opposed by a bulk interaction proportional to $R^2$ is described by an energy function
 \begin{eqnarray}
 U = \alpha R^2 c_2^2 + \beta R c_2
 \end{eqnarray}
 where $\alpha>0$ and $\beta$ are constants, giving $\bar c_2 = -\beta/2 \alpha R$. We can express the growing solutions of Eqn. 2 in a scaling form
 \begin{eqnarray}
 \frac{h_2^>}{R} =\frac{-\beta}{2 \alpha}  \left[ \left(\frac{r}{R} \right)^2 + \nu \left(\frac{r}{R} \right)^4 \right]
 \end{eqnarray}
 Thus for $m=2$ by expressing all lengths ($h,r$) in units of the disk radius $R$ one obtains a universal warped shape determined by the value of $\beta$. Note that this scaling rule is $m$-dependent, i.e. different $m$'s all show scaling but are described by different scaling functions. The full shape is scalable to that the extent it can be described by a single dominant angular harmonic. In Figure 3(b) we test this hypothesis by plotting the scaled height $h' = h_2^>/R$ versus the scaled radial coordinate $r'=r/R$ demonstrating its near collapse to a single profile. We conclude that an unwarped kirigami profile with no growing solutions is nongeneric, and would require fine tuning the system to a special point at $\beta =0$. This is evidently not a property of the LF potentials for carbon \cite{fasolino} nor of any generic model for the interparticle interactions.  Therefore the kirigami-ed disks generally feature a long distance shape modulation that cannot be confined to the defect. We interpret this as a microscopic analog to the step risers in macroscopic lattice kirigami that also propagate to the sample boundaries.  It also suggests the possibility of tuning the shape of such a system by functionalizing the system boundaries as mechanism for controlling the edge potential parameter $\beta$.

 \begin{figure}
\begin{center}
\includegraphics[angle=0,bb= 0 0 820 650,width=\columnwidth]{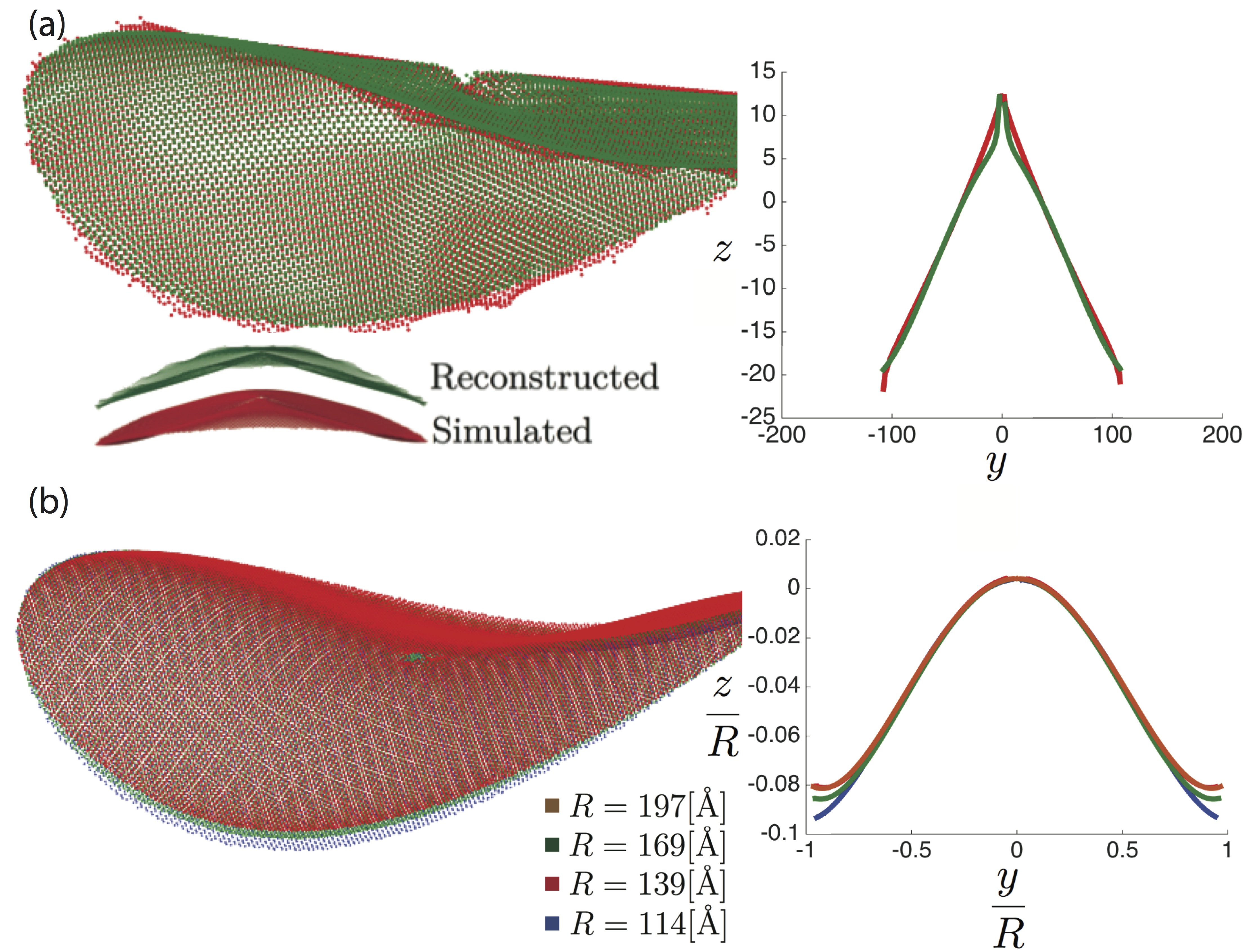}
\caption{\label{scaling} (a) Reconstruction of the {\it uu} surface retaining only the $m=0,\pm 2$ angular harmonics in the height field. (b) Numerical test of the scaling rule Eqn. 5 on four different disk radii. }
\end{center}
\end{figure}

 These considerations can also be used to understand the energetics of microscopic kirigami.  In macroscopic lattice kirigami the edges are sharp and the {\it uu} structure (Fig 1(c)) is degenerate in energy with the {\it ud}  structure (Fig. 1(d)). Furthermore the energy of an {\it uu} configuration is independent of the separation ($d$) of the dislocations that define the vertices of their plateaus (Fig. 1) since the sharp steps are nonoverlapping. These features do not apply to microscopic kirigami where the height profile is smooth and the dislocations can interact via overlap of their induced curvature fields. In Figure 4(a) we compare the energies of the {\it uu} and {\it ud} configurations as a function of the vertex separation $d$. (To obtain these data the relaxation calculations were carried out on square rather than circular models so that the number of atoms is the same in each sampled structure.)   The {\it uu} configuration is energetically preferred for any intervertex spacing $d$, though for large $d$ these energies converge to a common value which one can identify as twice the energy of a single dislocation. At intermediate separations the energy degeneracy is in fact strongly broken, for example the energy difference for a separation of $\sim 20 {\rm \AA}$ is $\approx 0.5$ eV.

 By analyzing these structures within continuum elastic theory we conclude that these energy differences arise from interactions that are mediated nearly entirely by the mean curvature of the extended overlapping height fields. The stretching energy, while present, is generally smaller than the energy stored in the mean curvature, and more importantly it is nearly $s$ independent, indicating that its role is to simply renormalize the total self energy in these structures. The interactions between defects mediated by the bending energy then lead qualitatively to the interaction profile shown in Figure 4. This behavior is captured even in a lowest order elastic theory. We first calculate the Lam${\rm \acute{e}}$ coefficients $\lambda$ and $\mu$ and bend modulus $\kappa_b$ using our model potential giving the values presented in Table 1. In this expansion the energy can be  partitioned into a pure bending contribution
 \begin{eqnarray}
 U_b = \frac{\kappa_b}{2} \, \int \, d^2 {\textbf{r}} \, (\nabla^2 h)^2
 \end{eqnarray}
 and a strain term
 \begin{eqnarray}
 U_s = \frac{1}{2} \, \int \, d^2 \textbf{r} \, (2 \mu u_{ij}^2 + \lambda u_{kk}^2 )
 \end{eqnarray}
 where $u_{ij}$ are the linearized strains $(\partial_i u_j + \partial_j u_i)/2$. (We have investigated the role of the nonlinear strain terms that can appear Eqn. 7 and find that they do not qualitatively change our conclusions.) Although the contribution from $U_s$ can be formally eliminated in favor of a (strongly) nonlocal interaction between Gaussian curvatures \cite{NelsonPeliti}, we choose instead to simply calculate the energy using the formula Eqn. 7.

 \begin{table}[h]
\centering
\begin{ruledtabular}
\begin{tabular}{c l}
Elastic constant & fitted value  \\
\hline
$\lambda$ & 3.23 ${\rm eV/\AA^2}$ \\
$\mu$ & 10.67 ${\rm eV/\AA^2}$ \\
$B$ & 12.29 ${\rm eV/\AA^2}$ \\
$\kappa_b$ & 93.49 ${\rm eV}$ \\
\end{tabular}
\end{ruledtabular}
\caption{\label{tab:5/tc}Two dimensional Lam${\rm \acute{e}}$ coefficients, bulk modulus and bending modulus obtained by fitting the structural energies for deformed graphene sheets using the interatomic potentials of Los and Fasolino \cite{fasolino}.}
\end{table}

  In the continuum model one finds that energy degeneracy of the {\it uu} and {\it ud} geometries is resolved and the {\it uu} configuration always favored. This can be understood if one regards the height fields of the two defects as additive. In the {\it uu} configuration the height deformations appear with opposite signs and nearly cancel in the far field while in the {\it ud} configuration they interfere constructively. The bending energy (though not the Gaussian curvature-induced stretching energy) is quadratic in derivatives of $h$ and so the relative signs of the superposed height fields determines the sign of the their interaction. This behavior captures the essential results of the full atomistic calculations (Fig. 4). We also note that while the results obtained from the bare elastic theory correctly describe the ordering of the structural energies it fails to quantitively account for their magnitudes, as can be expected since these structures are actually highly strained.

 \begin{figure}
\begin{center}
\includegraphics[angle=0,bb= 0 0 390 334,width=\columnwidth]{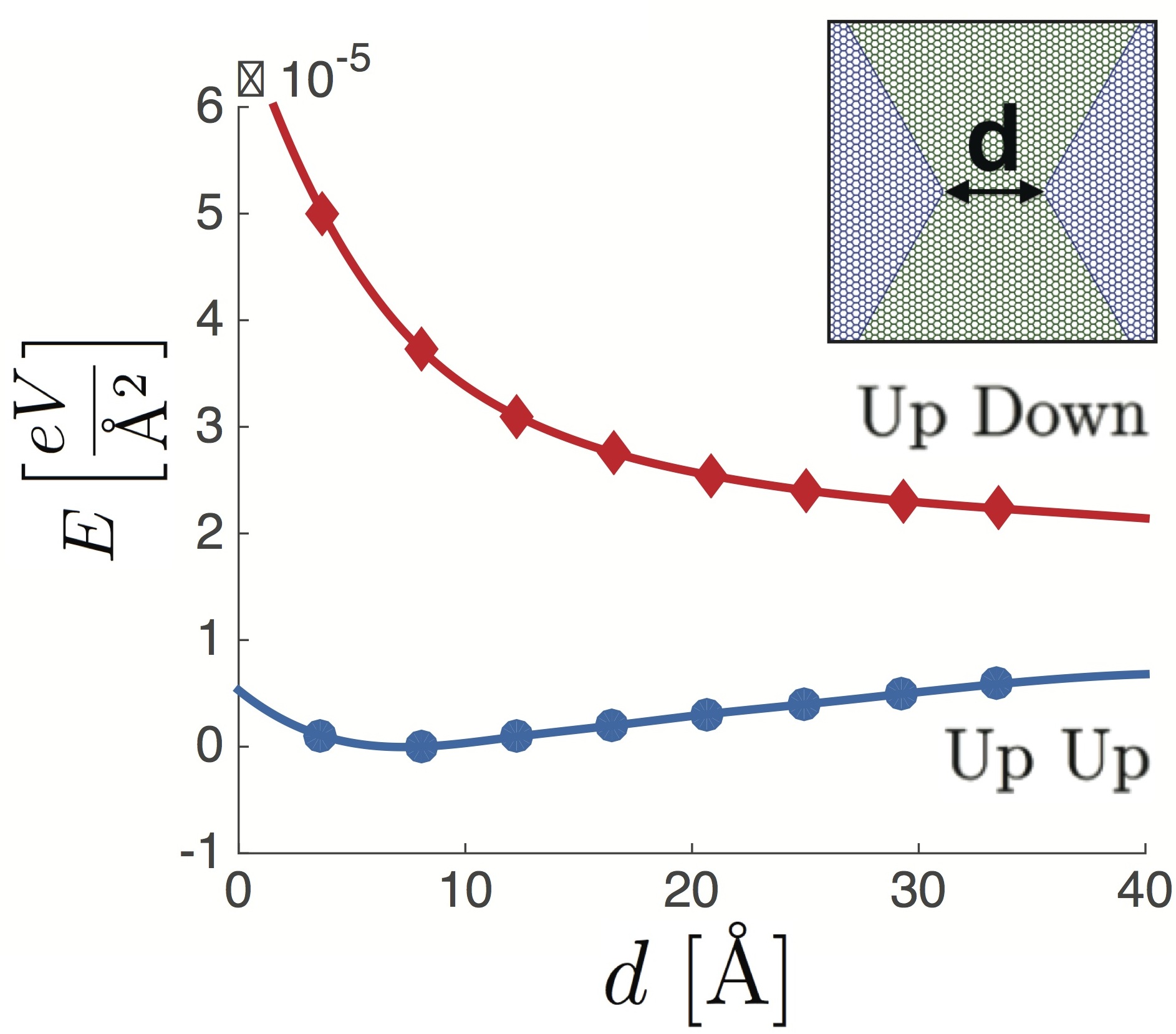}
\caption{\label{energies} Energies (expressed per unit area) for graphene kirigami as a function of vertex separation $d$. The {\it uu} and {\it ud} configurations are nondegenerate and the bend-induced {\it ud} potential is repulsive. These properties are described qualitatively within a continuum elastic theory where the energy differences and their $d$ dependence are controlled by the mean curvature in the relaxed structures.}
\end{center}
\end{figure}

 Insights from the bending energetics of nanoscale kirigami may be useful for stabilizing structures in macroscopic kirigami. The degeneracy of the {\it uu} and {\it ud} motifs is problematic for applications that would seek to stabilize a single target shape. This can be resolved by the introduction of macroscopic couplings that introduce an effective bending rigidity. Braces that suppress or promote bending can be engineered to introduce nonlocal coupling between neighboring step risers and provide a route to encoding a unique surface structure.

The analytic structure of our graphene-kirigami solutions also have important consequences for its Dirac electronic structure near charge neutrality. In these structures topological defects in their bond networks induce surface deformations with bend and (locally) nonzero Gaussian curvature. Separately, these structural features all couple to electronic motion in the tangent plane \cite{kanemele,zaanen,gaugefields,GaugedGraphene,phillips} where the natural language for this coupling involves valley asymmetric bend- and strain-induced gauge fields \cite{gaugefields}. The gauge fields induced by pure bend are curl-free and have the innocuous effect of simply shifting the Dirac points in momentum space. By contrast Gaussian curvature is topologically nontrivial and links the system with a (valley dependent) local flux \cite{GaugedGraphene}. The essential characteristic of the $m$-projected solutions presented above is that a competition between bending and stretching energies generates a landscape where the Gaussian curvature is globally compensated (so that the total pseudo-flux is zero) but this can only be accomplished by sign changes on a network of nodal lines that carry the signature of the fully relaxed kirigami.  The possibility of confining electronic modes along these lines and their role in defining the low energy spectral and transport properties now presents an important problem for further study.

This work was supported by the Department of Energy under grant DE FG02-84ER45118. EJM acknowledges support from the Leverhulme Trust at Loughborough University where this work was carried out.


\begin{thebibliography}{23}
\expandafter\ifx\csname natexlab\endcsname\relax\def\natexlab#1{#1}\fi
\expandafter\ifx\csname bibnamefont\endcsname\relax
  \def\bibnamefont#1{#1}\fi
\expandafter\ifx\csname bibfnamefont\endcsname\relax
  \def\bibfnamefont#1{#1}\fi
\expandafter\ifx\csname citenamefont\endcsname\relax
  \def\citenamefont#1{#1}\fi
\expandafter\ifx\csname url\endcsname\relax
  \def\url#1{\texttt{#1}}\fi
\expandafter\ifx\csname urlprefix\endcsname\relax\def\urlprefix{URL }\fi
\providecommand{\bibinfo}[2]{#2}
\providecommand{\eprint}[2][]{\url{#2}}

\bibitem{kirigami} T. Castle, Y. Cho, X. Gong, E. Jung, D. M. Sussman, S. Yang and R.D. Kamien, Physical Review Letters {\bf 113}, 245502 (2014).
\bibitem{YazyevReview} O.V. Yazyev and Y.P. Chen, Nature Nanotechnology {\bf 9}, 755 (2014).
    \bibitem{lammps} S. Plimpton, J. Comp.
Phys. {\bf 117}, 1-19 (1995) (available at http://lammps.sandia.gov).
\bibitem{fasolino} J. H. Los and A. Fasolino, Phys. Rev. B {\bf 68}, 024107 (2003).
\bibitem{LandauLifshitz} L.D. Landau and E.M. Lifshitz {\it Theory of Elasticity} (Pergamon, New York, 1970).
\bibitem{NelsonPeliti} D.R. Nelson and L. Peliti, J. Physique {\bf 48}, 108-1092 (1987).
\bibitem{SeungNelson} H.S. Seung and D.R. Nelson, Phys. Rev. A {\bf 38}, 1005 (1988).
\bibitem{graphenekirigamiBU} Z. Qi, D.K. Campbell and H.S. Park, Physical Review B {\bf 90}, 245437 (2014).
\bibitem{graphenekirigamiCornell} M. K. Blees, A. W. Barnard, P. A. Rose, S. P. Roberts, K. L.
McGill, P. Y. Huang, A. R. Ruyack, J. W. Kevek, B. Kobrin,
D. A. Muller, and P. L. McEuen, DOI:10.1038/nature14588
\bibitem{lehtinen} O. Lehtinen, S. Kurasch, A.V. Krasheninnikov and U. Kaiser, Nature Communications {\bf 4}, 2098 (2013).
\bibitem{kanemele} C.L. Kane and E.J. Mele, Physical Review Letters {\bf 78}, 1932 (1997).
\bibitem{zaanen} A. Mesaros, D. Sadri and J. Zaanen,  Physical Review B {\bf 79}, 155111 (2009).
\bibitem{GaugedGraphene} T.O. Wehling, A.V. Balatsky, A.M. Tsvelik, M.I. Katsnelson and A.I. Lichtenstein, Europhysics Letters {\bf 84}, 17003 (2008).
\bibitem{gaugefields} F. Guinea, M.I. Katsnelson and A.K. Geim, Nature Physics {\bf 6}, 30 (2010).
\bibitem{phillips} M. Phillips and E.J. Mele, Physical Review B {\bf 91}, 125404 (2015).
\end{thebibliography}
\end{document}